\documentclass[twocolumn,pre]{revtex4}

\usepackage{amsmath}
\usepackage{graphicx}

\newcommand{\e}{\mathrm{e}}
\newcommand{\eref}[1]{(\ref{#1})}
\newcommand{\set}[1]{\{#1\}}
\newcommand{\half}{\mbox{$\frac12$}}
\newcommand{\etal}{\textit{et al.}}
\newcommand{\av}[1]{\langle#1\rangle}
\newcommand{\bS}{S}
\newcommand{\ttheta}{\tilde{\theta}}
\newcommand{\talpha}{\tilde{\alpha}}
\newcommand{\bigav}[1]{\bigl\langle#1\bigr\rangle}
\newcommand{\defn}{\textit}

\begin{document}
\title{Solution for the properties of a clustered network}
\author{Juyong Park}
\affiliation{Department of Physics, University of Michigan, Ann Arbor,
  MI 48109--1120, U.S.A.}  
\author{M.~E.~J. Newman}
\affiliation{Department of Physics, University of Michigan, Ann Arbor,
  MI 48109--1120, U.S.A.} 
\begin{abstract}
We study Strauss's model of a network with clustering and present an
analytic mean-field solution which is exact in the limit of large network
size.  Previous computer simulations have revealed a degenerate region in
the model's parameter space in which triangles of adjacent edges clump
together to form unrealistically dense subgraphs, and perturbation
calculations have been found to break down in this region at all orders.
Our solution shows that this region corresponds to a classic
symmetry-broken phase and that the onset of the degeneracy corresponds to a
first-order phase transition in the density of the network.
\end{abstract}
\maketitle

\section{Introduction}
The last few years have seen a surge of interest within the scientific
community in the properties of networks of various
kinds~\cite{AB02,DM02,Newman03d}.  In parallel with empirical studies of
real-world networks such as the Internet~\cite{FFF99}, the worldwide
web~\cite{BA99b,Kleinberg99b}, biological networks~\cite{WM00,Ito01}, and
social networks~\cite{WS98}, researchers have developed theoretical models
and mathematical tools to explain the rich structure and nontrivial
characteristics that large-scale networks exhibit.

The most fundamental of network models may be the Bernoulli random
graph~\cite{Bollobas01} (also sometimes called the Erd\H{o}s--R\'enyi model
after two well-known mathematicians who were among the first to study
it~\cite{ER60}).  In this model, $n$~identical vertices are joined together
in pairs by edges, each possible edge appearing with independent
probability~$p$ for a total of ${n\choose2}p$ edges on average.  This model
can be thought of as a special case of the much larger class of
\defn{exponential random graphs}, which is the class of ensembles of graphs
that maximize ensemble entropy under a given set of constraints (usually
imposed by observations of the properties of an actual network in the real
world)~\cite{PN04b}.  The appropriate constraint for the Bernoulli random
graph is a constraint on the total number of edges in the graph.

The exponential random graph model defines a probability distribution over
a specified set of possible graphs such that the probability $P(G)$ of a
particular graph~$G$ is proportional to $\e^{-H(G)}$, where
\begin{equation}
H(G) = \sum_i \theta_i m_i(G).
\end{equation}
$H(G)$ is called the \defn{graph Hamiltionian}, $\set{m_i}$ is the set of
observables upon which the relevant constraints act, and $\set{\theta_i}$
is a set of real-valued conjugate fields which we can vary so as to match
the properties of the model to the real-world network under consideration.
Exact or approximate solutions of average properties of the ensemble are
possible for a variety of graph Hamiltonians, including graphs with
arbitrary degree distributions, directed-graph models with
reciprocity~\cite{PN04b}, the so-called 2-star model~\cite{PN04a}, and
others~\cite{PDFV04}.

In this paper we give a solution of a particular famous exponential random
graph model, the clustering model of Strauss~\cite{Strauss86}.  This model
mimics the phenomenon of network transitivity or clustering, which has been
much discussed in the networks literature~\cite{WS98,HK02b,KE02,Newman03e}.
The model was originally proposed in 1981 and has recently attracted the
attention of the physics community~\cite{BJK04a,BJK04b}, where the question
of how properly to model transitivity has proved a persistent stumbling
block for theorists.

\section{Strauss's model of clustering}
Strauss's model is simple to define.  The appropriate graph observables are
the number of edges~$m(G)$ and the number of triangles~$t(G)$, so that the
Hamiltonian can be written
\begin{eqnarray}
H(G) &=& \theta m(G) - \alpha t(G) \nonumber\\
     &=& \theta \sum_{i<j} \sigma_{ij} - \alpha \sum_{i<j<k}
                                       \sigma_{ij}\sigma_{jk}\sigma_{ki},
\label{hamil}
\end{eqnarray}
where $\sigma_{ij}=\sigma_{ji}$ is an element of the \defn{adjacency
matrix} having value $1$ if an edge exists between vertices $i$ and~$j$ and
$0$ otherwise.  When $\alpha>0$, this Hamiltonian encourages the formation
of triangles in the network by assigning lower ``energy'' to graphs with
many triangles.

Although the Hamiltonian seems simple enough, Strauss found via numerical
simulations that the model sometimes behaved strangely, developing in
certain parameter regimes a ``degenerate state,'' a condensed phase in
which many triangles form but tend to stick together in local regions of
the graph, rather than spreading uniformly over it.  Recently
Burda~\etal~\cite{BJK04a} have performed a perturbation theoretic analysis
of the model~\footnote{Our calculations differ from those of Burda~\etal\
in that Burda~\etal\ study sparse graphs with finite mean degree where we
study graphs with finite densities of edges.  However, the solution we give
should work in the sparse regime also, so direct comparison of the two
calculations is valid.}, finding that the formation of this condensed phase
corresponds to a point at which the perturbation series breaks down at all
orders simultaneously.  The nature of this point and of the condensed phase
however has not been well understood and a complete solution of the model
has been lacking.  In the next section, we present a solution of the model
based on a mean-field approach which we believe to be exact for all
parameter values in the limit of large system size.  Using this solution,
we show that the model possess a classic second-order phase transition
between a high-symmetry regime and a symmetry-broken one, with a line of
first-order transitions between states of high and low density in the
symmetry-broken regime.  The formation of the ``condensed phase'' observed
by Strauss corresponds precisely to the first-order transition from low to
high density.

\section{Analysis}
\subsection{Mean-field solution}
Let $H_{ij}$ be the sum of all terms in the Hamiltonian, Eq.~\eref{hamil},
that involve~$\sigma_{ij}$:
\begin{equation}
H_{ij} = \theta\sigma_{ij}
         - \alpha\sum_{k\ne i,j}\sigma_{ij}\sigma_{jk}\sigma_{ki}
       = \sigma_{ij} \bigl( \theta
                     - \alpha\sum_{k\ne i,j}\sigma_{jk}\sigma_{ki} \bigr),
\label{hij}
\end{equation}
and let $H'$ be the remaining terms, so that $H=H_{ij}+H'$.  The mean value
$\av{\sigma_{ij}}$ of $\sigma_{ij}$ can then be written as
\begin{eqnarray}
\av{\sigma_{ij}} &=& 0\times P(\sigma_{ij}=0)+1\times P(\sigma_{ij}=1)
\nonumber \\
                 &=& \frac{1}{Z}\sum_{\set{\sigma}} \e^{-H}
                     \frac{\e^{-H_{ij}(\sigma_{ij}=1)}}
                     {\e^{-H_{ij}(\sigma_{ij}=0)} + 
                     \e^{-H_{ij}(\sigma_{ij}=1)}} \nonumber\\
                 &=& \biggl\langle\frac{1}{\e^{\theta -
                     \alpha\sum_{k\ne i,j}\sigma_{jk}\sigma_{ki}}+1}
                     \biggr\rangle.
\label{exactsigma}
\end{eqnarray}
where $Z=\sum_G\e^{-H(G)}$ is the partition function.  Here $\av{\ldots}$
indicates the average within the ensemble, and the derivation so far has
been exact.

By analogy with spin models, let us call the expression within the brackets
in Eq.~\eref{hij} the \emph{local field} coupled to spin~$\sigma_{ij}$.
The mean-field approximation involves replacing the spin variables in the
local field with their ensemble averages, which in this case means
$\sigma_{jk}\sigma_{ki}\to q\equiv\av{\sigma_{jk}\sigma_{ki}}$.  Defining
also the
\textit{connectance} $p\equiv\av{\sigma_{ij}}$, we now have
\begin{equation}
p = \frac{1}{\e^{\theta-\alpha(n-2)q}+1} =
\half \bigl[ 1-\tanh \bigl( \half\theta-\half\alpha(n-2)q \bigr) \bigr].
\label{meansigma1}
\end{equation}

Now we set up an equation for $q$ via a similar method.
Noting that $\sigma_{ik}\sigma_{kj}=1$ only when both $\sigma_{ik}=1$
and $\sigma_{kj}=1$, we can write:
\begin{eqnarray}
q &\equiv& \av{\sigma_{ik}\sigma_{kj}} = \nonumber\\
  & & \hspace{-2em} \biggl\langle\frac{\e^{\alpha\sigma_{ij}}}
  {\bigl(\e^{\theta-\alpha\sum_l \sigma_{il}\sigma_{lk}}+1\bigr)
  \bigl(\e^{\theta-\alpha\sum_l\sigma_{kl}\sigma_{lj}}+1\bigr) + 
  (\e^{\alpha\sigma_{ij}}-1)}\biggr\rangle \nonumber\\
  &=&\frac{1+(\e^\alpha-1)p}{\bigl(\e^{\theta-\alpha(n-3)q}+1\bigr)^2
    + (\e^\alpha-1)p},
\label{meansigma2}
\end{eqnarray}
where in the final line we have made the mean-field approximation again,
and made use of the property that $\e^{\alpha\sigma_{ij}} =
1+(\e^\alpha-1)\sigma_{ij}$, since $\sigma_{ij}=0$ or~1.

We now have two equations in two unknowns which can be solved by
substituting~\eref{meansigma1} into~\eref{meansigma2} to give a
self-consistency condition on~$q$:
\begin{eqnarray}
q &=& \frac{\e^{\theta-\alpha (n-2) q}+\e^\alpha} {(\e^{\theta-\alpha
    (n-3) q}+1)^2 (\e^{\theta-\alpha (n-2) q}+1)+(\e^\alpha-1)} \nonumber\\
  &\equiv& Q(q).
\label{mft2}
\end{eqnarray}

\begin{figure}
\begin{center}
\resizebox{7cm}{!}{\includegraphics{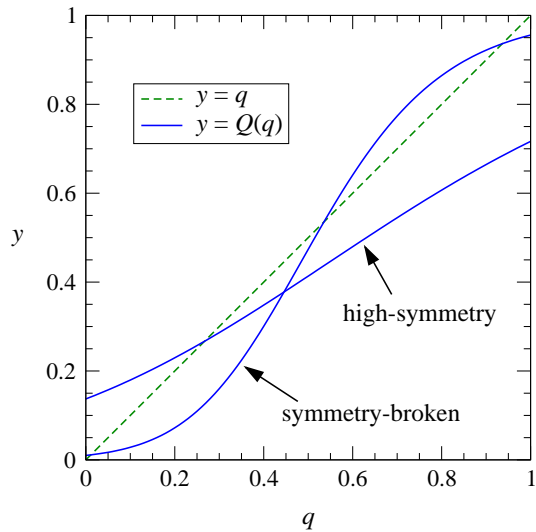}}
\end{center}
\caption{(Color online) Graphical solutions of $q=Q(q)$.  Depending on the
  values of the parameters $\theta$ and~$\alpha$, the line $y=q$ (dashed)
  intersects with $y=Q(q)$ (solid) either three times or only once.  The
  parameters $(\theta,n\alpha)$ for the two curves shown are $(2.3,6.0)$
  and $(0.5,2.0)$.}
\label{Qq}
\end{figure}

\begin{figure}
\begin{center}
\resizebox{7.5cm}{!}{\includegraphics{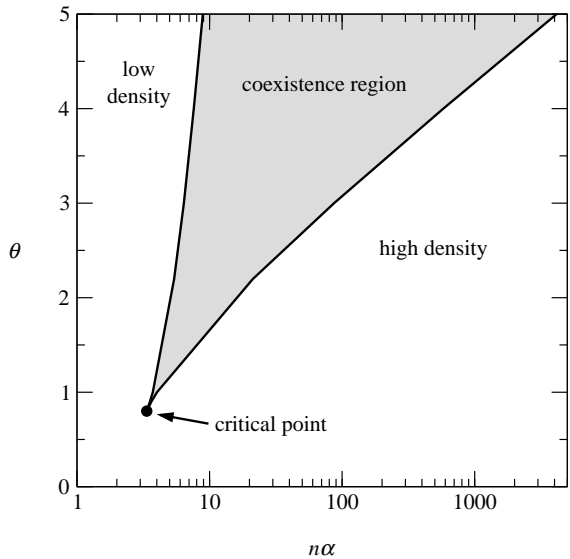}}
\end{center}
\caption{The phase diagram in the $(n\alpha,\theta)$ space.  The shaded
area corresponds to the coexistence region in which the system can be in
either of two stable states, one of high density and one of low.}
\label{phase}
\end{figure}

In Fig.~\ref{Qq} we show a plot of the forms $y=q$ and $y=Q(q)$ as
functions of~$q$.  The intersections of the two curves give the solutions
of Eq.~\eref{mft2}.  As we can see, depending on the values of $\theta$
and~$\alpha$, the curves can intersect at either one or three points in the
allowed domain $0<q<1$.  The regime in which there are three solutions
corresponds to a symmetric-broken phase with only the outer two solutions
being stable (corresponding to minima of the free energy).  Thus, the
system displays the classic phenomenology of a second-order phase
transition, with a critical point separating a high-symmetry phase from a
symmetry-broken one having regimes of high- and low-density and an
intermediate region of coexistence of the two.  In Fig.~\ref{phase} we show
the phase diagram of the system.

Finally we introduce another mean-field equation for
$r\equiv\av{\sigma_{ij}\sigma_{jk}\sigma_{ki}}$ which gives the number of
triangles in the network:
\begin{equation}
r \equiv \av{\sigma_{ij}\sigma_{jk}\sigma_{ki}}
  = \frac{\e^\alpha}{\bigl(\e^{\theta-\alpha(n-2)q}+1\bigr)^3+(\e^\alpha-1)}.
\label{meansigma3}
\end{equation}

\begin{figure}
\begin{center}
\resizebox{8.5cm}{!}{\includegraphics{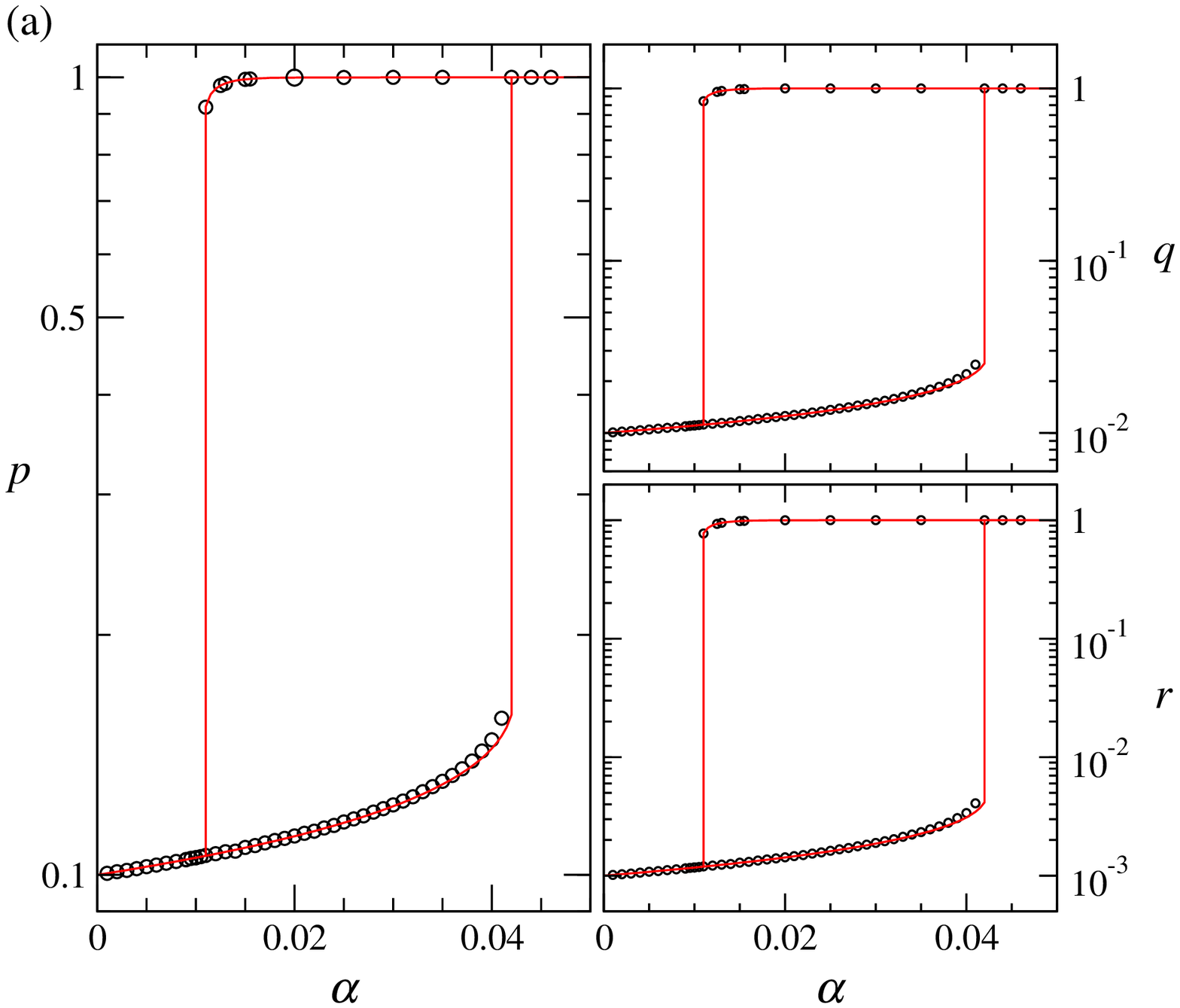}}
\resizebox{8.5cm}{!}{\includegraphics{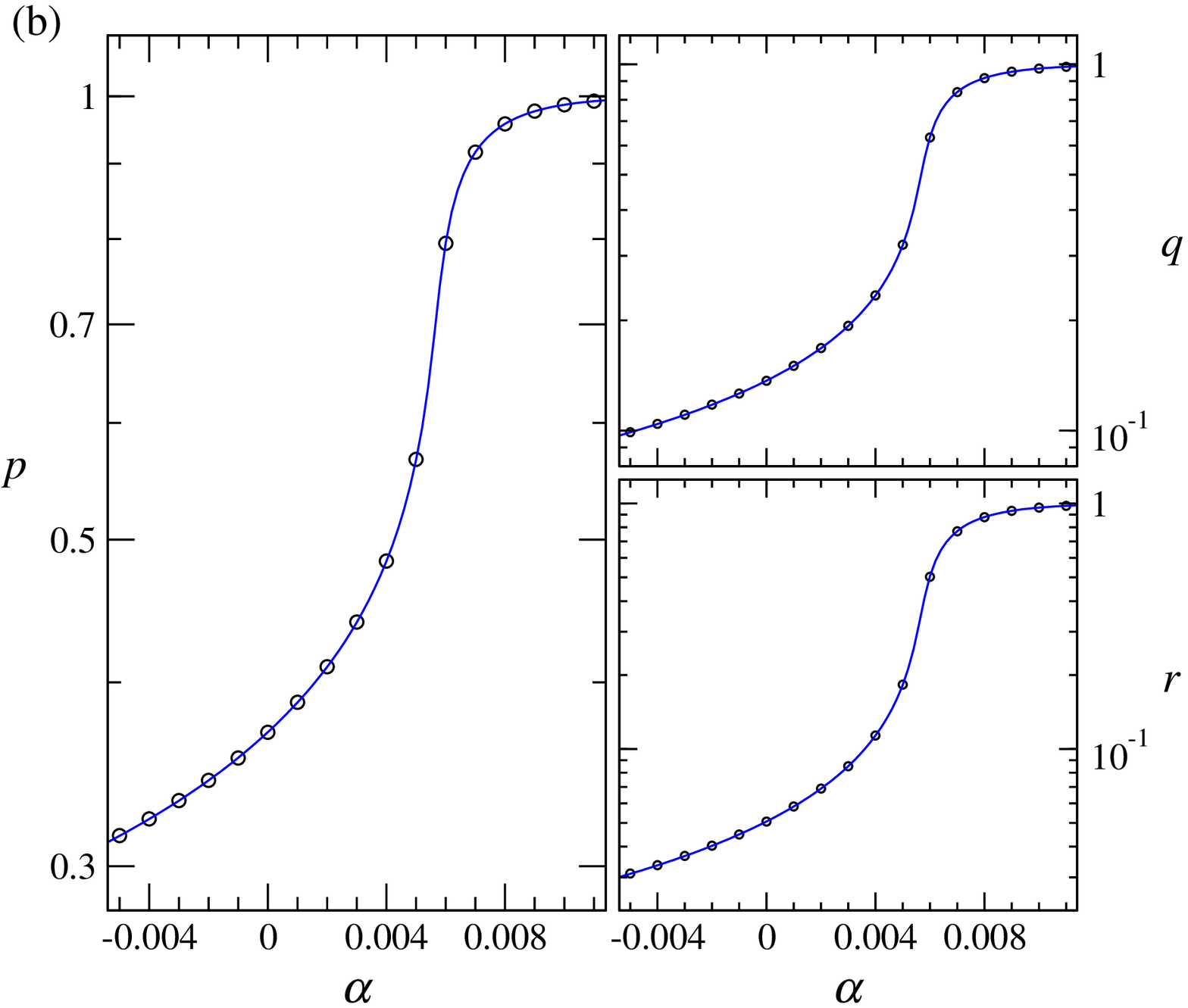}}
\end{center}
\caption{(Color online) Comparison of our analytic solution (solid lines)
and Monte Carlo simulation results (circles) for $p=\av{\sigma_{ij}}$,
$q=\av{\sigma_{jk}\sigma_{ki}}$, and
$r=\av{\sigma_{ij}\sigma_{jk}\sigma_{ki}}$, for a system of $n=500$
vertices.  The parameter values were (a)~$\theta=2.2$ and
(b)~$\theta=0.53$.  (See Fig.~\ref{phase}.)}
\label{plot}
\end{figure}

In Fig.~\ref{plot} we compare our solutions for $p$, $q$ and~$r$ with
simulation results for a system of size $n=500$ and, as we can see, the
agreement between theory and simulation is excellent.  

\subsection{The approximation}
As mentioned in the introduction, we believe that the mean-field solution
found in the previous section is exact, because in the limit of large
system size the system becomes fundamentally infinite-dimensional, and
mean-field theory is usually exact in the large dimension limit.

In fact, Eq.~\eref{exactsigma} really makes two approximations.  One is the
mean-field approximation
$\sigma_{jk}\sigma_{ki}\to\av{\sigma_{jk}\sigma_{ki}}$, but we have also
assumed that the average of the tanh can be approximated by the tanh of the
average.  While the first approximation can be justified on the basis of
the high effective dimension of the system, the second needs more
attention.  It can be justified by performing a series expansion of the
tanh, applying the mean-field approximation to the series term by term, and
then resumming the result again~\cite{Parisi88,BCM91}.  However, while this
method works, it is not as simple as our brief description makes it sound,
because the series involves averages over arbitrarily high moments of the
graph operators and proving that these terms are negligible requires some
care.

Let us rewrite Eq.~\eref{exactsigma} thus:
\begin{eqnarray}
p \equiv \av{\sigma_{ij}} &=& \half \bigl( 1 + \bigl\langle
  \tanh (-\half\theta+\half\alpha\sum_{k\ne i,j} S_k) \bigr\rangle \bigr)
  \nonumber\\
&=& \half \bigl( 1 + \bigl\langle
\tanh\bigl(\tilde{\theta}+\tilde{\alpha}\bS\bigr) \bigr\rangle \bigr),
\label{mft1}
\end{eqnarray}
where for convenience we have defined $\tilde{\theta}=-\half\theta$,
$\tilde{\alpha}=\half\alpha$, $S_k=\sigma_{ik}\sigma_{kj}$, and
$\bS=\sum_{k\ne i,j}S_k$.  Expanding the tanh about $\ttheta$, we get
\begin{equation}
\bigl\langle\tanh(\ttheta+\talpha\bS)\bigr\rangle 
= \sum_{m=0}^{\infty}\frac{\tanh^{(m)}(\ttheta)\talpha^m}{m!} \bigav{\bS^m}.
\label{hstaylor}
\end{equation}
Now, keeping in mind that
$S_k^m=\sigma_{ik}^m\sigma_{kj}^m=\sigma_{ik}\sigma_{kj}=S_k$ for any~$m$,
since $\sigma_{ij}=0,1$, we can write the correlation functions $\av{S^m}$
in the form
\begin{subequations}
\begin{eqnarray}
\label{hsmfta}
\av{\bS^m} &=&\bigav{\bigl(\mbox{$\sum_k$}S_k\bigr)
  \bigl(\mbox{$\sum_k$}S_k\bigr)\ldots \bigl(\mbox{$\sum_k$}S_k\bigr)} \\ 
  & & \hspace{-2em} = a_{m,1}\bigav{S_1+S_2+\ldots} 
      + a_{m,2}\bigav{S_1S_2+S_1S_3+\ldots} \nonumber\\
  & & {} + a_{m,3}\bigav{S_1S_2S_3+\ldots} +\ldots \\
  & & \hspace{-2em} = a_{m,1}\mbox{$(n-2)$}q +
      a_{m,2}\mbox{$\binom{n-2}{2}$}q^2+a_{m,3}\mbox{${n-2\choose 
      3}$}q^3+\ldots\nonumber\\
\label{hsmftc}
\end{eqnarray}
\label{hsmft}
\end{subequations}
In the last line we have made the assumption that a product
$\bigav{S_1S_2}$ can be approximated as $\bigav{S_1}\bigav{S_2}=q^2$, and
similarly for higher products.  This approximation is of the nature of a
mean-field approximation, ignoring correlations between single pairs of
spin variables, which will be of order $1/n$ in the large system size limit
where each variable interacts with an arbitrary number of others.

The coefficient $a_{m,l}$ in Eq.~\eref{hsmft} is the number of ways of
selecting one $S_k$ from each of the $m$ sums in~\eref{hsmfta} so that
there are $l$ unique indices in the resulting product.  It is simple to
show that $\sum_{i=1}^{l}\binom{l}{i} a_{m,i}=l^m$ and thus by induction to
prove that the exponential generating function for $a_{m,l}$ satisfies
\begin{equation}
g_l(z) = \sum_{m=1}^{\infty} {z^m a_{m,l}\over m!} = (\e^z-1)^l.
\end{equation}
Then, by repeated differentiation
\begin{equation}
a_{m,l} = \biggl[ \frac{\partial^m~}{\partial z^m}(\e^z-1)^l\biggr]_{z=0}. 
\end{equation}
Combining this result with Eq.~\eref{hsmft} and taking the limit of
large~$n$, we find
\begin{eqnarray}
\av{\bS^m} &=& \sum_{l=0}^{\infty}a_{m,l}\mbox{${n-2\choose l}$}q^l
           \simeq \sum_{l=0}^{\infty}a_{m,l} (n-2)^lq^l/l! \nonumber\\
           &=& \biggl[ \frac{\partial^m~}{\partial z^m} \sum_{l=0}^{\infty}
               \frac{(\e^z-1)^l(n-2)^lq^l}{l!}\biggl]_{z=0} \nonumber\\
           &=& \biggl[ \frac{\partial^m~}{\partial z^m} 
               \e^{(n-2)q(\e^z-1)} \biggl]_{z=0}.
\end{eqnarray}

The differentiation in the last line can be carried out explicitly for any
given value of~$m$, but there is no simple closed-form expression for the
general case.  However, none is needed in the large~$n$ limit.  Each
successive differentiation with respect to~$z$ generates an extra factor of
$(n-2)q$.  But the graphs we are interested in are \textit{dense}, meaning
that $p$, $q$, and $r$ all tend to finite, non-zero limiting values as
$n\to\infty$.  Thus $(n-2)q$ is a large quantity and to leading order we
need only retain the highest-order term in the derivative, which is simply
$\bigl[(n-2)q\bigr]^m$.  Thus Eq.~\eref{hstaylor} becomes
\begin{eqnarray}
\bigl\langle\tanh(\ttheta+\talpha\bS)\bigr\rangle
   &=& \sum_{m=0}^{\infty}\frac{\tanh^{(m)}(\ttheta)\talpha^m}{m!}
\bigav{\bS^m} \nonumber\\
     &=& \sum_{m=0}^{\infty} \frac{\tanh^{(m)}(\ttheta)\talpha^m}{m!}
\bigl((n-2)q\bigr)^m \nonumber\\
     &=& \tanh\bigl(\ttheta+\talpha (n-2)q\bigr)
\end{eqnarray}
and
\begin{equation}
p = \half \bigl[ 1 + \tanh\bigl(\ttheta+\talpha (n-2)q\bigr) \bigr],
\end{equation}
which is identical with Eq.~\eref{meansigma1}.  A similar derivation can be
performed for Eq.~\eref{meansigma2}, and hence the entire mean-field
solution is exact in the limit of large system size.

\section{Discussion}
What does our solution of the Strauss model tell us?  To begin with, it
tells us the precise nature of and reason for the ``degenerate state''
observed by Strauss in simulation studies and by Burda~\etal~\cite{BJK04a}
in their perturbative calculations.  Strauss's observations were
correct---something special does happen to the model in the degenerate
region.  In fact, there is a first-order phase transition driven by the
``field'' parameter coupled to the number of edges in the graph.  This also
explains the breakdown of the perturbation expansion at this point, since
such expansions typically break down at first-order transitions because of
the corresponding pole in the free energy.  The degenerate phase of the
model is a high-density phase in which there is a large number of triangles
in the graph, forming what appears to be almost a complete graph: the
connectance of the network is close to~1 in this regime (Fig.~\ref{plot}).

More importantly, the first-order nature of the transition means there is a
discontinuous jump in the density of triangles as we enter the degenerate
state and thus there is no intermediate set of parameter values that will
give the graph a moderate density of triangles as seen in real-world
networks.  While Strauss's model seems the most natural form for an
exponential random graph model of transitivity, our results imply that it
will in fact never be a good model of real-world networks with moderate
clustering.  One can of course reduce the value of the parameter~$\alpha$
until we pass through the critical point so that the first-order transition
disappears, in which case we recover smooth variation of the density of the
graph with~$\theta$, but then the graph no longer has any significant
clustering because of the small value of~$\alpha$.

These observations do not necessarily imply that exponential random graphs
are incapable of mimicking networks with clustering; indeed they may
present our best current hope for making clustered network models.  Our
results imply however that Strauss's original model with a single term in
the Hamiltonian to encourage triangles must, at the very least, be
augmented in some way in order to achieve this aim.

\section{Conclusions}
In this paper we have given a mean-field solution of Strauss's model of a
network with clustering.  Because of the intrinsically high-dimensional
nature of networks, we believe this solution to be exact in the limit of
large system size, which is the main case one is normally interested in.
We have also performed Monte Carlo simulations of the model that confirm
our solution to high accuracy.  Our solution indicates that the model has
no regime in which it displays moderate levels of clustering similar to
those seen in real-world networks; presumably it will be necessary to
introduce further terms into the Hamiltonian to avoid this pathology.

We believe exponential random graphs offer one of the most flexible tools
for the modeling of general networks, and look forward to further
developments.  We hope that the formalism introduced here will serve as a
practical starting point for a variety of problems.

\begin{acknowledgments}
This work was funded in part by the National Science Foundation under grant
number DMS--0405348.
\end{acknowledgments}


\begin{thebibliography}{22}
\expandafter\ifx\csname natexlab\endcsname\relax\def\natexlab#1{#1}\fi
\expandafter\ifx\csname bibnamefont\endcsname\relax
  \def\bibnamefont#1{#1}\fi
\expandafter\ifx\csname bibfnamefont\endcsname\relax
  \def\bibfnamefont#1{#1}\fi
\expandafter\ifx\csname citenamefont\endcsname\relax
  \def\citenamefont#1{#1}\fi
\expandafter\ifx\csname url\endcsname\relax
  \def\url#1{\texttt{#1}}\fi
\expandafter\ifx\csname urlprefix\endcsname\relax\def\urlprefix{URL }\fi
\providecommand{\bibinfo}[2]{#2}
\providecommand{\eprint}[2][]{\url{#2}}

\bibitem[{\citenamefont{Albert and Barab\'asi}(2002)}]{AB02}
\bibinfo{author}{\bibfnamefont{R.}~\bibnamefont{Albert}} \bibnamefont{and}
  \bibinfo{author}{\bibfnamefont{A.-L.} \bibnamefont{Barab\'asi}},
  \bibinfo{journal}{Rev. Mod. Phys.} \textbf{\bibinfo{volume}{74}},
  \bibinfo{pages}{47} (\bibinfo{year}{2002}).

\bibitem[{\citenamefont{Dorogovtsev and Mendes}(2002)}]{DM02}
\bibinfo{author}{\bibfnamefont{S.~N.} \bibnamefont{Dorogovtsev}}
  \bibnamefont{and} \bibinfo{author}{\bibfnamefont{J.~F.~F.}
  \bibnamefont{Mendes}}, \bibinfo{journal}{Advances in Physics}
  \textbf{\bibinfo{volume}{51}}, \bibinfo{pages}{1079} (\bibinfo{year}{2002}).

\bibitem[{\citenamefont{Newman}(2003{\natexlab{a}})}]{Newman03d}
\bibinfo{author}{\bibfnamefont{M.~E.~J.} \bibnamefont{Newman}},
  \bibinfo{journal}{SIAM Review} \textbf{\bibinfo{volume}{45}},
  \bibinfo{pages}{167} (\bibinfo{year}{2003}{\natexlab{a}}).

\bibitem[{\citenamefont{Faloutsos et~al.}(1999)\citenamefont{Faloutsos,
  Faloutsos, and Faloutsos}}]{FFF99}
\bibinfo{author}{\bibfnamefont{M.}~\bibnamefont{Faloutsos}},
  \bibinfo{author}{\bibfnamefont{P.}~\bibnamefont{Faloutsos}},
  \bibnamefont{and}
  \bibinfo{author}{\bibfnamefont{C.}~\bibnamefont{Faloutsos}},
  \bibinfo{journal}{Computer Communications Review}
  \textbf{\bibinfo{volume}{29}}, \bibinfo{pages}{251} (\bibinfo{year}{1999}).

\bibitem[{\citenamefont{Barab\'asi and Albert}(1999)}]{BA99b}
\bibinfo{author}{\bibfnamefont{A.-L.} \bibnamefont{Barab\'asi}}
  \bibnamefont{and} \bibinfo{author}{\bibfnamefont{R.}~\bibnamefont{Albert}},
  \bibinfo{journal}{Science} \textbf{\bibinfo{volume}{286}},
  \bibinfo{pages}{509} (\bibinfo{year}{1999}).

\bibitem[{\citenamefont{Kleinberg et~al.}(1999)\citenamefont{Kleinberg, Kumar,
  Raghavan, Rajagopalan, and Tomkins}}]{Kleinberg99b}
\bibinfo{author}{\bibfnamefont{J.~M.} \bibnamefont{Kleinberg}},
  \bibinfo{author}{\bibfnamefont{S.~R.} \bibnamefont{Kumar}},
  \bibinfo{author}{\bibfnamefont{P.}~\bibnamefont{Raghavan}},
  \bibinfo{author}{\bibfnamefont{S.}~\bibnamefont{Rajagopalan}},
  \bibnamefont{and} \bibinfo{author}{\bibfnamefont{A.}~\bibnamefont{Tomkins}},
  in \emph{\bibinfo{booktitle}{Proceedings of the International Conference on
  Combinatorics and Computing}} (\bibinfo{publisher}{Springer},
  \bibinfo{address}{Berlin}, \bibinfo{year}{1999}), no. \bibinfo{number}{1627}
  in \bibinfo{series}{Lecture Notes in Computer Science}, pp.
  \bibinfo{pages}{1--18}.

\bibitem[{\citenamefont{Williams and Martinez}(2000)}]{WM00}
\bibinfo{author}{\bibfnamefont{R.~J.} \bibnamefont{Williams}} \bibnamefont{and}
  \bibinfo{author}{\bibfnamefont{N.~D.} \bibnamefont{Martinez}},
  \bibinfo{journal}{Nature} \textbf{\bibinfo{volume}{404}},
  \bibinfo{pages}{180} (\bibinfo{year}{2000}).

\bibitem[{\citenamefont{Ito et~al.}(2001)\citenamefont{Ito, Chiba, Ozawa,
  Yoshida, Hattori, and Sakaki}}]{Ito01}
\bibinfo{author}{\bibfnamefont{T.}~\bibnamefont{Ito}},
  \bibinfo{author}{\bibfnamefont{T.}~\bibnamefont{Chiba}},
  \bibinfo{author}{\bibfnamefont{R.}~\bibnamefont{Ozawa}},
  \bibinfo{author}{\bibfnamefont{M.}~\bibnamefont{Yoshida}},
  \bibinfo{author}{\bibfnamefont{M.}~\bibnamefont{Hattori}}, \bibnamefont{and}
  \bibinfo{author}{\bibfnamefont{Y.}~\bibnamefont{Sakaki}},
  \bibinfo{journal}{Proc. Natl. Acad. Sci. USA} \textbf{\bibinfo{volume}{98}},
  \bibinfo{pages}{4569} (\bibinfo{year}{2001}).

\bibitem[{\citenamefont{Watts and Strogatz}(1998)}]{WS98}
\bibinfo{author}{\bibfnamefont{D.~J.} \bibnamefont{Watts}} \bibnamefont{and}
  \bibinfo{author}{\bibfnamefont{S.~H.} \bibnamefont{Strogatz}},
  \bibinfo{journal}{Nature} \textbf{\bibinfo{volume}{393}},
  \bibinfo{pages}{440} (\bibinfo{year}{1998}).

\bibitem[{\citenamefont{Bollob\'as}(2001)}]{Bollobas01}
\bibinfo{author}{\bibfnamefont{B.}~\bibnamefont{Bollob\'as}},
  \emph{\bibinfo{title}{Random Graphs}} (\bibinfo{publisher}{Academic Press},
  \bibinfo{address}{New York}, \bibinfo{year}{2001}), \bibinfo{edition}{2nd}
  ed.

\bibitem[{\citenamefont{Erd\H{o}s and R\'enyi}(1960)}]{ER60}
\bibinfo{author}{\bibfnamefont{P.}~\bibnamefont{Erd\H{o}s}} \bibnamefont{and}
  \bibinfo{author}{\bibfnamefont{A.}~\bibnamefont{R\'enyi}},
  \bibinfo{journal}{Publications of the Mathematical Institute of the Hungarian
  Academy of Sciences} \textbf{\bibinfo{volume}{5}}, \bibinfo{pages}{17}
  (\bibinfo{year}{1960}).

\bibitem[{\citenamefont{Park and Newman}(2004{\natexlab{a}})}]{PN04b}
\bibinfo{author}{\bibfnamefont{J.}~\bibnamefont{Park}} \bibnamefont{and}
  \bibinfo{author}{\bibfnamefont{M.~E.~J.} \bibnamefont{Newman}},
  \bibinfo{journal}{Phys. Rev. E} \textbf{\bibinfo{volume}{70}},
  \bibinfo{pages}{066117} (\bibinfo{year}{2004}{\natexlab{a}}).

\bibitem[{\citenamefont{Park and Newman}(2004{\natexlab{b}})}]{PN04a}
\bibinfo{author}{\bibfnamefont{J.}~\bibnamefont{Park}} \bibnamefont{and}
  \bibinfo{author}{\bibfnamefont{M.~E.~J.} \bibnamefont{Newman}},
  \bibinfo{journal}{Phys. Rev. E} \textbf{\bibinfo{volume}{70}},
  \bibinfo{pages}{066146} (\bibinfo{year}{2004}{\natexlab{b}}).

\bibitem[{\citenamefont{Palla et~al.}(2004)\citenamefont{Palla, Der\'enyi,
  Farkas, and Vicsek}}]{PDFV04}
\bibinfo{author}{\bibfnamefont{G.}~\bibnamefont{Palla}},
  \bibinfo{author}{\bibfnamefont{I.}~\bibnamefont{Der\'enyi}},
  \bibinfo{author}{\bibfnamefont{I.}~\bibnamefont{Farkas}}, \bibnamefont{and}
  \bibinfo{author}{\bibfnamefont{T.}~\bibnamefont{Vicsek}},
  \bibinfo{journal}{Phys. Rev. E} \textbf{\bibinfo{volume}{69}},
  \bibinfo{pages}{046117} (\bibinfo{year}{2004}).

\bibitem[{\citenamefont{Strauss}(1986)}]{Strauss86}
\bibinfo{author}{\bibfnamefont{D.}~\bibnamefont{Strauss}},
  \bibinfo{journal}{SIAM Review} \textbf{\bibinfo{volume}{28}},
  \bibinfo{pages}{513} (\bibinfo{year}{1986}).

\bibitem[{\citenamefont{Holme and Kim}(2002)}]{HK02b}
\bibinfo{author}{\bibfnamefont{P.}~\bibnamefont{Holme}} \bibnamefont{and}
  \bibinfo{author}{\bibfnamefont{B.~J.} \bibnamefont{Kim}},
  \bibinfo{journal}{Phys. Rev. E} \textbf{\bibinfo{volume}{65}},
  \bibinfo{pages}{026107} (\bibinfo{year}{2002}).

\bibitem[{\citenamefont{Klemm and Eguiluz}(2002)}]{KE02}
\bibinfo{author}{\bibfnamefont{K.}~\bibnamefont{Klemm}} \bibnamefont{and}
  \bibinfo{author}{\bibfnamefont{V.~M.} \bibnamefont{Eguiluz}},
  \bibinfo{journal}{Phys. Rev. E} \textbf{\bibinfo{volume}{65}},
  \bibinfo{pages}{036123} (\bibinfo{year}{2002}).

\bibitem[{\citenamefont{Newman}(2003{\natexlab{b}})}]{Newman03e}
\bibinfo{author}{\bibfnamefont{M.~E.~J.} \bibnamefont{Newman}},
  \bibinfo{journal}{Phys. Rev. E} \textbf{\bibinfo{volume}{68}},
  \bibinfo{pages}{026121} (\bibinfo{year}{2003}{\natexlab{b}}).

\bibitem[{\citenamefont{Burda et~al.}(2004{\natexlab{a}})\citenamefont{Burda,
  Jurkiewicz, and Krzywicki}}]{BJK04a}
\bibinfo{author}{\bibfnamefont{Z.}~\bibnamefont{Burda}},
  \bibinfo{author}{\bibfnamefont{J.}~\bibnamefont{Jurkiewicz}},
  \bibnamefont{and}
  \bibinfo{author}{\bibfnamefont{A.}~\bibnamefont{Krzywicki}},
  \bibinfo{journal}{Phys. Rev. E} \textbf{\bibinfo{volume}{69}},
  \bibinfo{pages}{026106} (\bibinfo{year}{2004}{\natexlab{a}}).

\bibitem[{\citenamefont{Burda et~al.}(2004{\natexlab{b}})\citenamefont{Burda,
  Jurkiewicz, and Krzywicki}}]{BJK04b}
\bibinfo{author}{\bibfnamefont{Z.}~\bibnamefont{Burda}},
  \bibinfo{author}{\bibfnamefont{J.}~\bibnamefont{Jurkiewicz}},
  \bibnamefont{and}
  \bibinfo{author}{\bibfnamefont{A.}~\bibnamefont{Krzywicki}},
  \bibinfo{journal}{Phys. Rev. E} \textbf{\bibinfo{volume}{70}},
  \bibinfo{pages}{026106} (\bibinfo{year}{2004}{\natexlab{b}}).

\bibitem[{\citenamefont{Parisi}(1988)}]{Parisi88}
\bibinfo{author}{\bibfnamefont{G.}~\bibnamefont{Parisi}},
  \emph{\bibinfo{title}{Statistical Field Theory}}
  (\bibinfo{publisher}{Addison-Wesley}, \bibinfo{address}{Reading, MA},
  \bibinfo{year}{1988}).

\bibitem[{\citenamefont{Banavar et~al.}(1991)\citenamefont{Banavar, Cieplak,
  and Maritan}}]{BCM91}
\bibinfo{author}{\bibfnamefont{J.~R.} \bibnamefont{Banavar}},
  \bibinfo{author}{\bibfnamefont{M.}~\bibnamefont{Cieplak}}, \bibnamefont{and}
  \bibinfo{author}{\bibfnamefont{A.}~\bibnamefont{Maritan}},
  \bibinfo{journal}{Phys. Rev. Lett.} \textbf{\bibinfo{volume}{67}},
  \bibinfo{pages}{1807} (\bibinfo{year}{1991}).

\end{thebibliography}
\end{document}